\documentclass{article}

\usepackage{arxiv}

\usepackage[utf8]{inputenc} 
\usepackage[T1]{fontenc}    
\usepackage{hyperref}       
\usepackage{url}            
\usepackage{booktabs}       
\usepackage{amsfonts}       
\usepackage{nicefrac}       
\usepackage{microtype}      
\usepackage{lipsum}
\usepackage{times}
\usepackage{epsfig}
\usepackage{graphicx}
\usepackage{amsmath}
\usepackage{amssymb}
\usepackage{caption,setspace}
\usepackage{subfig}
\usepackage{cite}
\usepackage{CJK}
\usepackage{float}
\usepackage{color}
\usepackage{verbatim}
\usepackage{indentfirst}
\title{Sampled-Data Control Based Consensus of Fractional-Order Multi-Agent Systems}

\author{
  Xinyao~Li \\
    School of Electrical and Electronic Engineering\\
Nanyang Technological University\\
Singapore, 639798\\
\texttt{E180209@e.ntu.edu.sg}\\
   \And
 Changyun~Wen \\
  School of Electrical and Electronic Engineering\\
Nanyang Technological University\\
Singapore, 639798\\
  \texttt{ecywen@ntu.edu.sg} \\
  \And
  Xiao-Kang~Liu \\
  School of Electrical and Electronic Engineering\\
Nanyang Technological University\\
Singapore, 639798\\
\texttt{xiaokang.liu@ntu.edu.sg} \\
}

\begin{document}
\maketitle

\begin{abstract}
In this paper, we investigate consensus control of fractional-order multi-agent systems with order in $(0,1)$ via sampled-data control. A new scheme to design distributed controllers with rigorous analysis is presented by utilizing the unique properties of fractional-order calculus, namely hereditary and infinite memory. It is established that global boundedness of all closed-loop signals is ensured and asymptotic consensus is realized. Simulation studies are conducted to illustrate the effectiveness of the proposed control method and verify the obtained results.
\end{abstract}


\section{Introduction}
Multi-agent systems are widely studied in the past decades because of their applications in various areas such as formation control, distributed sensor network, and so on \cite{fax2004information, cortes2005coordination, yu2009distributed}. Generally, the target of distributed consensus control is to achieve an agreement for the states of all the systems connected in a network by designing a suitable controller for each agent depending on locally available information from itself and its neighbors. 

Recently, consensus control problems for integer-order multi-agent systems have been broadly investigated, see for examples \cite{olfati2004consensus, wang2014distributed, wang2017distributed, cheng2011necessary, cheng2016convergence, huang2015adaptive, huang2017smooth, chen2009reaching, zheng2018average}, and references therein. 
As for the consensus control problem of fractional-order multi-agent systems, it is first studied in \cite{cao2009distributed}, where the conditions for achieving consensus in term of network structure and the number of agents are provided. In \cite{sun2011convergence}, where an undirected graph is considered, a consensus protocol with switching order is raised to increase convergence speed. Consensus involving communication delays is addressed in \cite{shen2012necessary}. Observer-based leader-following consensus problem is investigated in \cite{yu2016observer}, where the leader is described as a second-order integer model while the followers are fractional-order systems with order $\alpha \in(0,2)$. Consensus for incommensurate nonlinear fractional-order multi-agent systems with system uncertainties and external disturbances is addressed in \cite{shahvali2019distributed}.

All the works above are studied with continuous control, which requires control signals to be updated and transmitted continuously. In comparison with continuous control, sampled-data control for continuous-time systems, which can be found in for example \cite{hu2007sampled} and \cite{shen2016stabilization}, possesses various of benefits such as low cost and be more practical in implementation. 
Many research works for integer-order multi-agent systems with sampled-data control have also been done and a survey on this topic is provided in \cite{ge2018survey}. However, results on sampled-data control of fractional-order multi-agent systems are still limited. The consensus problem of such systems with directed graph via sampled-data control is investigated in \cite{yu2017necessary, su2019necessary, liu2019consensus}. In \cite{chen2019synchronization}, consensus of linear fractional-order multi-agent systems over a communication topology, whose coupling structure is not necessary to be Laplacian, is studied. Event-triggering sampled-data control of fractional-order multi-agent systems is proposed in \cite{chen2019consensus}, in which the networked graph for the agents is assumed to be undirected. 

From the definitions of fractional integral and derivative reviewed in Section \ref{sec:Preliminaries}, it can be observed that they can both be treated as weighted integral, which reveals the properties of fractional-order calculus: hereditary and infinite memory. Due to these unique properties, the initial values and the ``history" of the variables in the entire interval of integration play extremely important roles in solving fractional-order equations. It is noted that such properties are not considered in the existing literature on sampled-data control of fractional-order multi-agent systems mentioned above. Instead, by dividing the entire time history to sampling intervals, the solutions of fractional-order equations at the beginning of each sampling interval are used as the new initial values for this sampling interval. As a result, the closed-loop fractional-order
systems obtained are transformed to discrete-time systems, where the evolutions of step-forward states only rely on the current states and control inputs. However, when taking these properties into account, the evolutions should also depend on the previous states at all the discrete-time sampling instants starting from the initial time $t_0$. 

Motivated by the discussions above, we address the consensus issue with the consideration of such unique characteristics of fractional-order calculus. With our proposed control design scheme, it is shown that the closed-loop system is globally stable and asymptotic consensus for fractional-order multi-agent systems is achieved. Different from existing research works, 
the resulting fractional-order system is rigorously analyzed by considering time evolution of the system states which depends on their initial values at $t_0$ and also all the previous control signals. A new challenge that is to establish the global boundedness of the proposed distributed controllers which contain all the previous control inputs is overcome in this paper. Simulation studies illustrate the effectiveness of the proposed control scheme and also reveal its accuracy on achieving asymptotic consensus compared to an existing approach. 

The rest of the paper is organized as follows. Preliminaries are provided and the class of  fractional-order multi-agent systems considered is described in Section \ref{sec:Preliminaries}. In Section \ref{sec:controller}, the design of distributed controllers is presented in detail with analysis. In Section \ref{sec:simulation}, the scheme is illustrated by simulation studies with comparison to that in \cite{yu2017necessary}. Finally the paper is concluded in Section \ref{sec:conclusion}.

\section{Preliminaries and Problem Formulation}
\label{sec:Preliminaries}


\subsection{Preliminaries}
\emph{Definition 1 \cite{podlubny1998fractional}:} The fractional integral of an integrable function $f(t)$ with $\alpha\in\mathbb{R}^{+}$ and initial time $t_0$ is
\begin{equation}
{}_{t_0}\!{\mathit{I}}_t^{\alpha}f(t) = \frac{1}{\Gamma(\alpha)}\int_{t_0}^{t}\frac{f(\tau)}{(t-\tau)^{1-\alpha}}\,d\tau
\label{eq:eqI}
\end{equation}
where $\Gamma(\cdot)$ denotes the well-known Gamma function, which is defined as $\Gamma(z)=\int_{0}^{\infty}e^{-t}t^{z-1}\,dt$, where $z \in \mathbb{C}$. One of the significant properties of Gamma function is \cite{xue2017fractional}: $\Gamma(z+1) = z\Gamma(z),\, \Gamma(n+1) = n\Gamma(n)=n(n-1)\Gamma(n-1)=\cdots=n!,\, \Gamma(-n)= \infty$, where $n \in \mathbb{N}_0=\{n|n \ge 0,n \in \mathbb{N}\}$.

\emph{Definition 2 \cite{podlubny1998fractional}:} The Caputo fractional derivative of a function is defined as
\begin{equation}
\begin{aligned}
{}_{t_0}^{C}\!{\mathcal{D}}_t^{\alpha}f(t) &= {}_{t_0}{\mathit{I}}_t^{(m-\alpha)}\frac{\mathrm{d}^m}{\mathrm{d}t^m}f(t)\\
&=\frac{1}{\Gamma(m-\alpha)}\int_{t_0}^{t} \frac{f^{(m)}(\tau)}{(t-\tau)^{\alpha-m+1}}\,d\tau\\
\end{aligned}
\label{eq:eqCd}
\end{equation}
where $m - 1 < \alpha < m \in\mathbb{Z}^+$. From (\ref{eq:eqCd}) we can observe that the Caputo derivative of a constant is $0$.  Another commonly used fractional derivative is named Riemann-Liouville (RL) and the RL fractional derivative of a function $f(t)$ is denoted as ${}_{t_0}^{RL}\!{\mathcal{D}}_t^{\alpha}f(t)$. Different from the Caputo derivative, RL derivative of a constant is not equal to $0$ \cite{lakshmikantham2009theory,podlubny1998fractional}.

The initial values are needed in order to obtain the unique solution for fractional differential equation ${}_{t_0}\!{\mathcal{D}}_t^{\alpha}x(t)=f(x,t)$, ($m - 1 < \alpha < m \in\mathbb{Z}^+$ and $t\ge t_0$). According to \cite{li2007remarks,podlubny1998fractional} and \cite{bandyopadhyay2015stabilization}, fractional differential equations with Caputo-type derivative have initial values that are in-line with integer-order differential equations, i.e.\ $x(t_0),x'(t_0),\dots,x^{(m-1)}(t_0)$, which contain specific physical interpretations. 
Therefore, Caputo-type fractional systems are frequently employed in practical analysis.

\emph{Definition 3 \cite{li2009mittag,li2010stability, zhang2011asymptotical}:} For fractional nonautonomous system ${}_{t_0}\!{\mathcal{D}}_t^{\alpha}x_i(t)=f_i(x,t)$ ($i = 1,2,\dots,n$), where $0 < \alpha < 1$, initial condition $x(t_0) = [x_1(t_0),x_2(t_0),\dots,x_n(t_0)]^\mathrm{T} \in \mathbb{R}^n$, $f_i(x,t): [t_0, \infty) \times \Omega \rightarrow \mathbb{R}^n$ is locally Lipschitz in $x$ and piecewise continuous in $t$ (which insinuates the existence and uniqueness of the solution to the fractional systems \cite{podlubny1998fractional}), ${}_{t_0}\!{\mathcal{D}}_t^{\alpha}$ denotes Caputo or RL fractional derivative and $\Omega \in \mathbb{R}^n$ stands for a region that contains the origin $x = [0,0,\dots,0]^\mathrm{T}$, the equilibrium $x^* =[x_1^*,x_2^*,\dots,x_n^*]^\mathrm{T}$ of this system is defined as ${}_{t_0}\!{\mathcal{D}}_t^{\alpha}x^*=f_i(x^*,t)$ for $t \ge t_0$. 

\emph{Lemma 1 \cite{lakshmikantham2009theory}:} If $x(t)\in \mathbb{R}^n$ satisfies
\begin{equation}
\begin{aligned}
{}_{t_0}^{C}\!{\mathcal{D}}_t^{\alpha}x(t) &= f(x,t), \; x(t_0)=x_0,
\end{aligned}
\label{eq:eqL1_1}
\end{equation}
where $0<\alpha<1$ and $f(x,t) \in L_1[t_0,T]$, then it also satisfies the Volterra fractional integral
\begin{equation}
\begin{aligned}
x(t)=x_0+\frac{1}{\Gamma(\alpha)}\int_{t_0}^t (t-\tau)^{\alpha-1}f\big(x(\tau),\tau\big)d\tau,
\end{aligned}
\label{eq:eqL1_2}
\end{equation}
with $t_0\leq t \leq T$ and vice versa.

\emph{Lemma 2:} For $\forall j \in\mathbb{Z}^+$ and $0<\alpha<1$, the following results hold\\
 \indent1) $0<\lvert (j+1)^\alpha-2j^\alpha+(j-1)^\alpha\rvert <1$,\\
 \indent2) $\lim_{j \rightarrow \infty} \lvert (j+1)^\alpha -2j^\alpha+(j-1)^\alpha\rvert = 0$.

\emph{Proof:} Define 
\begin{equation}
\begin{aligned}
f(j)=(j+1)^\alpha -2j^\alpha+(j-1)^\alpha
\end{aligned}
\label{eq:eqL2_1}
\end{equation}
and 
\begin{equation}
\begin{aligned}
g(s)&=s^\alpha-(s-1)^\alpha.
\end{aligned}
\label{eq:eqL2_g}
\end{equation}
It can be shown from (\ref{eq:eqL2_g}) that, for $\forall s \in\mathbb{R}^+$,
\begin{equation}
\begin{aligned}
g(s)&> 0,\\
\frac{dg(s)}{ds}=\alpha s^{\alpha-1}&-\alpha (s-1)^{\alpha-1} < 0.
\end{aligned}
\label{eq:eqL2_2}
\end{equation}
From (\ref{eq:eqL2_2}), we can obtain that $g(s)$ is monotonically decreasing and $\lim_{s\rightarrow \infty}g(s)=0$, which further implies that $f(j)<0$ and will monotonically tends to 0 as $j$ tends to $\infty$. Additionally, it can be easily checked that $\lvert f(1)\rvert < 1$ for $0<\alpha<1$. This complete the proof.

\subsection{Problem Formulation}
In this paper, Caputo-type definition of the fractional derivatives is utilized. A group of $\mathit{N}$ fractional-order agents are governed by
\begin{equation}
\begin{aligned}
{}_{t_0}^{C}\!{\mathcal{D}}_t^{\alpha}&x_{i}(t) = u_i(t), \text{ for } i=1,2,\cdots,N
\end{aligned}
\label{eq:eqPro}
\end{equation} 
where the fractional-orders of all the states are equal to $\alpha \in (0, 1)$,
$x_i\in \mathbb{R}$ and $u_i \in \mathbb{R}$ represent the measurable state and control input of $i$-th agent, respectively.

\emph{Remark 1:} All the agents in this paper are in one-dimensional space for convenience. The results established can be easily extended to $n$-dimensional space by applying the Kronecker product.

In this paper, the control problem is to design distributed controller $u_i$ for each agent described in (\ref{eq:eqPro}) to achieve the following objectives: 1) all the signals in the closed-loop systems are globally bounded; 2) asymptotic consensus for fractional-order systems (\ref{eq:eqPro}) is ensured, i.e.\ $\lim_{t\rightarrow \infty}\lVert x_i(t)-x_j(t)\rVert=0,\, \forall i, j = 1,2,\cdots,\mathit{N}$, and additionally, $\lim_{{t \rightarrow \infty}}x_i(t)=\frac{\sum_{i=1}^N x_i(t_0)}{N}$.

Suppose that the communications among the $\mathit{N}$ agents can be represented by a directed graph $\mathcal{G}\triangleq (\mathcal{V},\mathcal{E})$ where $\mathcal{V}=\{1,2,\cdots,\mathit{N}\}$ means the set of indexes (or vertices) corresponding to each agent, $\mathcal{E} \subseteq \mathcal{V} \times \mathcal{V}$ is the set of edges between two distinct agents. An edge $(i,j) \in \mathcal{E}$ denotes that agent $j$ can obtain information from agent $i$, but not necessarily vice versa. In this case, agent $i$ is called a neighbor of agent $j$ and we indicate the set of neighbors for agent $i$ as $\mathcal{N}_i$. In this paper, $(i,i) \notin \mathcal{E}$ and $i\notin \mathcal{N}_i$ since self edges $(i,i)$ are not allowed. $\mathit{A}=[a_{ij}]\in \mathbb{R}^{N\times N}$ is the connectivity matrix with $a_{ij}=1$ if $(j,i)\in \mathcal{E}$ and $a_{ij}=0$ if $(j,i)\notin \mathcal{E}$ defined. Throughout this paper, the diagonal elements $a_{ii}=0$. An in-degree matrix $\triangle$ is introduced as $\triangle = \text{diag}(\triangle_i)\in \mathbb{R}^{N \times N}$ with $\triangle_i=\sum_{j\in \mathcal{N}_i}a_{ij}$ being the $i$-th row sum of $\mathit{A}$. Therefore, the Laplacian matrix of $\mathcal{G}$ is defined as $\mathcal{L}=\triangle-\mathit{A}$. A digraph is strongly connected if there is a directed path that connects any two arbitrary nodes of the graph and is balanced if for all $i \in \mathcal{V}$, $\sum_{j \ne i}a_{ij}=\sum_{j \ne i}a_{ji}$. 

\emph{Notations:} $\lVert \cdot \rVert$ is the Euclidean norm of a vector. $I_N$ denotes an identity matrix with dimension equals to $N$. $\mathbf{1}_N=[1,1,\cdots,1]^{\mathrm{T}}\in \mathbb{R}^N$. 

\emph{Assumption 1:} The digraph $\mathcal{G}$ is strongly connected and balanced.

\section{Distributed Controller Design and Stability Analysis}
\label{sec:controller}
\subsection{Distributed Controller Design}
To achieve the above objectives, distributed controller is designed for each local agent based on periodic sampled-data control technology. The sampling instants are described by a discrete-time sequence $\{t_k\}$ with $t_0 < t_1 < \cdots < t_k < \cdots$ and $t_{k+1}-t_k = h$, where $h>0$ is the sampling period.

For $t\in [t_0, t_1)$, let $x(t)=[x_1(t),x_2(t),\cdots,x_N(t)]^{\mathrm{T}}$ and $\underline{u}_0=[u_{1,0},u_{2,0},\cdots,u_{N,0}]^{\mathrm{T}}$ where $u_{i,0}$ denotes the control signal for the $i$-th agent within this time interval. According to Lemma 1, the value of $x(t_1)$ can be computed as follows
\begin{equation}
\begin{aligned}
x(t_1)&=x(t_0)+\frac{1}{\Gamma(\alpha)}\int_{t_0}^{t_1}(t_1-\tau)^{\alpha-1}\underline{u}_0d\tau\\
&=x(t_0)+\underline{u}_0\frac{h^\alpha}{\Gamma(\alpha+1)}.
\end{aligned}
\label{eq:eqxi1_1}
\end{equation}

By designing the controller for the $i$-th agent as
\begin{equation}
u_{i,0}=\gamma \sum_{j\in \mathcal{N}_i}a_{ij}\big(x_j(t_0)-x_i(t_0)\big),
\label{eq:equ0i}
\end{equation}
where $\gamma > 0$, the closed-loop systems in this sampling period can be expressed as
\begin{equation}
\begin{aligned}
x(t_1)&=x(t_0)-\gamma\mathcal{L}x(t_0)\frac{h^\alpha}{\Gamma(\alpha+1)}.
\end{aligned}
\label{eq:eqxi1_2}
\end{equation}

Now for $t\in [t_k, t_{k+1})\,(k=1,2,\cdots)$, define $\underline{u}_k=[u_{1,k},u_{2,k},\cdots,u_{N,k}]^{\mathrm{T}}$. Since the fractional-order derivative of $x$ depends on all the historical values of $x$, thus not only the values of $x$ at present instant but also all of its previous values are needed to determine the future behavior of fractional-order systems. Therefore, $x(t_{k+1})$ should be expressed in terms of $x(t_0)$ and $\underline{u}_j (j=0,1,\cdots,k)$ as follows
\begin{equation}
\begin{aligned}
x(t_{k+1})=&x(t_0)+\frac{1}{\Gamma(\alpha)}\int_{t_0}^{t_1}(t_{k+1}-\tau)^{\alpha-1}\underline{u}_0d\tau+\cdots+\frac{1}{\Gamma(\alpha)}\int_{t_k}^{t_{k+1}}(t_{k+1}-\tau)^{\alpha-1}\underline{u}_k d\tau\\
=&x(t_k)+\sum_{l=1}^k\frac{[(l+1)h]^\alpha-2(lh)^\alpha+[(l-1)h]^\alpha}{\Gamma(\alpha+1)}\underline{u}_{k-l}+\frac{\underline{u}_k}{\Gamma(\alpha+1)}h^\alpha.
\end{aligned}
\label{eq:eqxik_1}
\end{equation}
Then based on (\ref{eq:eqxik_1}), we design the distributed controller for the $i$-th agent as
\begin{equation}
\begin{aligned}
u_{i,k}=&\gamma \sum_{j\in\mathcal{N}_i}a_{ij}\big(x_j(t_k)-x_i(t_k)\big)-\sum_{l=1}^{k}[(l+1)^\alpha-2l^\alpha+(l-1)^\alpha]u_{i,(k-l)},
\end{aligned}
\label{eq:equki}
\end{equation}
which results in the following closed-loop systems
\begin{equation}
\begin{aligned}
x(t_{k+1})=&x(t_k)-\gamma\mathcal{L}x(t_k)\frac{h^\alpha}{\Gamma(\alpha+1)}.
\end{aligned}
\label{eq:eqxik_2}
\end{equation}

\emph{Remark 2:} Existing studies on sampled-data control of fractional-order multi-agent systems divide the whole time interval $[t_0,t]$ into sampling intervals $[t_k,t_{k+1}],\,k=0,1,\cdots$ and solve the fractional-order equations on each sampling interval by treating $x(t_k)$ as the initial value for this corresponding time period. Through this way, the closed-loop fractional-order systems are simply modeled by simplified discrete-time systems in such a way that $x(t_{k+1})$ is expressed as a function only of $x(t_k)$ and $\underline{u}_k$, as is done in the integer-order derivative systems discretization. However, due to the hereditary and infinite memory properties of fractional-order derivative, when converting the actual initial value $x(t_0)$ into $x(t_k)$ for solving fractional-order equations, the second term $\sum_{l=1}^k\frac{[(l+1)h]^\alpha-2(lh)^\alpha+[(l-1)h]^\alpha}{\Gamma(\alpha+1)}\underline{u}_{k-l}$ on the right-hand side of (\ref{eq:eqxik_1}) cannot be neglected. On the contrary, the control scheme design and system analysis in this paper are carried out strictly by bearing the unique properties of fractional-order calculus in mind, which specifically can be seen from (\ref{eq:eqxik_1}), (\ref{eq:equki}) and the boundedness analysis of control signals given later.

\emph{Remark 3:} Note that the second term on the right-hand side of (\ref{eq:eqxik_1}) depends on all the previous control signals $\underline{u}_j,\,j=0,1,\cdots,k-1,$ and thus it cannot be assumed bounded before establishing system stability, giving arise to a challenge in controller design and analysis. Such a challenge is overcome in our proposed controller in (\ref{eq:equki}) which consists of two parts. The first part $u_{i1,k}=\gamma \sum_{j\in\mathcal{N}_i}a_{ij}\big(x_j(t_k)-x_i(t_k)\big) (k=0,1,2,\cdots)$ is designed for achieving consensus of fractional-order multi-agent systems and the second part $u_{i2,k}=-\sum_{l=1}^{k}[(l+1)^\alpha-2l^\alpha+(l-1)^\alpha]u_{i,(k-l)}$ which exists for $t \geq t_1$ aims at compensating for the effect caused by the hereditary and infinite memory properties of fractional-order calculus. It can be observed from (\ref{eq:eqxi1_2}) and (\ref{eq:eqxik_2}) that our designed distributed controllers allow us to analyze the stability of the closed-loop systems in the same way as that of the integer-order discrete-time closed-loop systems, which will be demonstrated in the next subsection. 

\subsection{Stability Analysis}
Our main result is presented in the following theorem, where a stability criterion is given. 

\emph{Theorem 1:}  Consider the closed-loop systems consisting of fractional systems (\ref{eq:eqPro}) and sampled-data based distributed controllers (\ref{eq:equ0i}) and (\ref{eq:equki}). All the signals in the closed-loop systems are globally bounded and asymptotic consensus is achieved, i.e.\ $\lim_{t\rightarrow \infty}\lVert x_i(t)-x_j(t)\rVert=0,\, \forall i, j = 1,2,\cdots,\mathit{N}$, if the design parameters $h$ and $\gamma$ satisfy
\begin{equation}
0 < \gamma\frac{h^\alpha}{\Gamma(\alpha+1)} <\frac{1}{\triangle_{\text{max}}}
\label{eq:eqh}
\end{equation}
and
\begin{equation}
\bigg[\beta^{k+1}+\sum_{l=1}^{k+1}\lvert f(l)\rvert \bigg] \leq 1,
\label{eq:eqT2}
\end{equation}
in which $\triangle_{\text{max}}$ is the maximum degree of the graph $\mathcal{G}$, $f(l)$ is defined in (\ref{eq:eqL2_1}) and
\begin{equation}
\beta = 1-\gamma \lambda_2(\mathcal{L}_s)\frac{h^\alpha}{\Gamma(\alpha+1)},
\label{eq:eqbeta}
\end{equation}
where $\mathcal{L}_s=\frac{\mathcal{L}+\mathcal{L}^\mathrm{T}}{2}$ and $\lambda_2(\mathcal{L}_s)$ represents the second smallest eigenvalue of $\mathcal{L}_s$. Moreover, since the digraph is balanced, asymptotic average-consensus can be achieved, i.e. $\lim_{{t \rightarrow \infty}}x_i(t)=\frac{\sum_{i=1}^N x_i(t_0)}{N}$.

\emph{Proof:} As mentioned above, the main challenge is how to achieve global stability of the resulting systems in the presence of the second term on the right-hand side of (\ref{eq:eqxik_1}). For this purpose, an additional control action is proposed in (\ref{eq:equki}), in order to compensate for the effects of this term. However, with this new control term which is the weighted sum of previous control signals, it is difficult to show the boundedness of control inputs. To overcome this difficulty, we first establish the following relationship
\begin{equation}
 \lVert \underline{u}_k\rVert  \leq \lVert \underline{u}_0\rVert ,\,\text{for all}\, k \geq 0.
\label{eq:eqL3}
\end{equation}

Now we define error vectors as
\begin{equation}
e(t_k)=\mathcal{L}x(t_k),\, k=0,1,\cdots.
\label{eq:eqek}
\end{equation}
Then the proof of (\ref{eq:eqL3}) is completed through mathematical induction as detailed below.

\emph{Step 1:} According to (\ref{eq:equ0i}), (\ref{eq:equki}) and (\ref{eq:eqek}), $\underline{u}_0$ and $\underline{u}_1$ can be respectively expressed as follows 
\begin{equation}
\underline{u}_0 = -\gamma e(t_0),
\label{eq:equ0}
\end{equation}
\begin{equation}
\underline{u}_1 = -\gamma e(t_1) + \lvert f(1) \rvert \underline{u}_0.
\label{eq:equ1}
\end{equation}

Therefore, we can have
\begin{equation}
\begin{aligned}
\lVert \underline{u}_1\rVert  =& \big\lVert -\gamma e(t_1) + \lvert f(1) \rvert \underline{u}_0 \big\rVert \\
\leq &  \gamma \lVert e(t_1) \rVert +\lvert f(1) \rvert \cdot\lVert \underline{u}_0 \rVert\\
\leq&  \gamma \beta\lVert e(t_0) \rVert +\lvert f(1) \rvert \cdot\lVert \underline{u}_0 \rVert \\
= & \big[ \beta +\lvert f(1) \rvert \big]\cdot  \lVert \underline{u}_0 \rVert.
\end{aligned}
\label{eq:equ1u0}
\end{equation}

By designing $\gamma$ and $h$ in such a way that $\big[\beta + \lvert f(1)\rvert\big] \leq 1$, then we can have $ \lVert \underline{u}_1\rVert  \leq \lVert \underline{u}_0\rVert $.

\emph{Step 2:} Assuming that $ \lVert \underline{u}_k\rVert  \leq \lVert \underline{u}_0\rVert $ holds for $k>1$. According to (\ref{eq:equki}) and (\ref{eq:eqek}), $ \lVert \underline{u}_{k+1}\rVert $ can be expressed as
\begin{equation}
\begin{aligned}
\lVert \underline{u}_{k+1}\rVert = & \bigg\lVert-\gamma e(t_{k+1})+\sum_{l=1}^{k+1}\lvert f(l) \rvert \underline{u}_{k+1-l} \bigg\rVert  \\
\leq &  \gamma \lVert e(t_{k+1}) \rVert+\sum_{l=1}^{k+1}\lvert f(l) \rvert \cdot \lVert \underline{u}_{k+1-l}\rVert \\
\leq & \gamma \beta^{k+1}\lVert e(t_0)\rVert +\sum_{l=1}^{k+1}\lvert f(l)\rvert \cdot\lVert \underline{u}_{k+1-l}\rVert .
\end{aligned}
\label{eq:equk+1u0_1}
\end{equation}
Since for $1 \leq l \leq k+1$, $ \lVert \underline{u}_{k+1-l}\rVert \leq \lVert \underline{u}_{0}\rVert $ holds, hence 
\begin{equation}
\begin{aligned}
\lVert \underline{u}_{k+1}\rVert  \leq & \beta^{k+1}\lVert \underline{u}_0\rVert +\sum_{l=1}^{k+1}\lvert f(l)\rvert \cdot\lVert \underline{u}_{0}\rVert \\
=& \bigg[ \beta^{k+1}+\sum_{l=1}^{k+1} \lvert f(l) \rvert \bigg]\cdot\lVert \underline{u}_{0}\rVert.
\end{aligned}
\label{eq:equk+1u0_2}
\end{equation}

If $\gamma$ and $h$ are chosen to satisfy (\ref{eq:eqT2}), then we have (\ref{eq:eqL3}).

Since $\lVert\underline{u}_0 \rVert= \gamma \lVert e(t_0)\rVert$ is bounded, therefore the global boundedness of all control signals are guaranteed.

Furthermore, (\ref{eq:eqxik_2}) can be rewritten as
\begin{equation}
\begin{aligned}
x(t_{k+1})=Px(t_k)
\end{aligned}
\label{eq:eqxik_3}
\end{equation}
where $P=I_N-\gamma\frac{h^\alpha}{\Gamma(\alpha+1)}\mathcal{L}$ is Perron matrix of digraph $\mathcal{G}$ with $\gamma\frac{h^{\alpha}}{\Gamma(\alpha+1)}$ treated as an integrated gain. Under condition (\ref{eq:eqh}), all the eigenvalues of $P$ are within the unit circle, which further implies the global boundedness of $x(t)$.

Also, from (\ref{eq:eqxik_2}) it can be noticed that the $i$-th agent in the closed-loop systems is described as 
\begin{equation}
x_i(t_{k+1})=x_i(t_k)+\gamma\frac{h^{\alpha}}{\Gamma(\alpha+1)}\sum_{j\in\mathcal{N}_i}a_{ij}\big(x_j(t_k)-x_i(t_k)\big).
\label{eq:eqTh1}
\end{equation}
 Hence, under Assumption 1, the proof of the condition (\ref{eq:eqh}) for reaching asymptotic consensus follows from that of Theorem 2 in \cite{olfati2007consensus} and thus the theorem is proved.

\section{Illustrative Example}
\label{sec:simulation}
In this section, an example is presented to demonstrate the proposed design control scheme and verify the established theoretical results. By comparing to an existing scheme in \cite{yu2017necessary}, it is revealed that the proposed controller can achieve asymptotic consensus in a more precise way.

Consider a group of five fractional-order agents with the following dynamics
\begin{equation}
\begin{aligned}
{}_{t_0}^{C}\!{\mathcal{D}}_t^{\alpha}&x_{i}(t) = u_i(t), \, i=1,2,\cdots,5,
\end{aligned}
\label{eq:eqsim}
\end{equation} 
where $\alpha=0.9$ and initial values of states are $x(t_0)=[4.5,5,6,1.5,-1]^{\mathrm{T}}$. 
The connection weights of the graph are $a_{14}=a_{15}=a_{21}=a_{23}=a_{31}=a_{32}=a_{42}=a_{43}=a_{54}=1$ and other entries of $\mathit{A}$ are equal to zero.

\begin{figure}
\centering
\includegraphics[width=10.4cm,height=6.4cm]{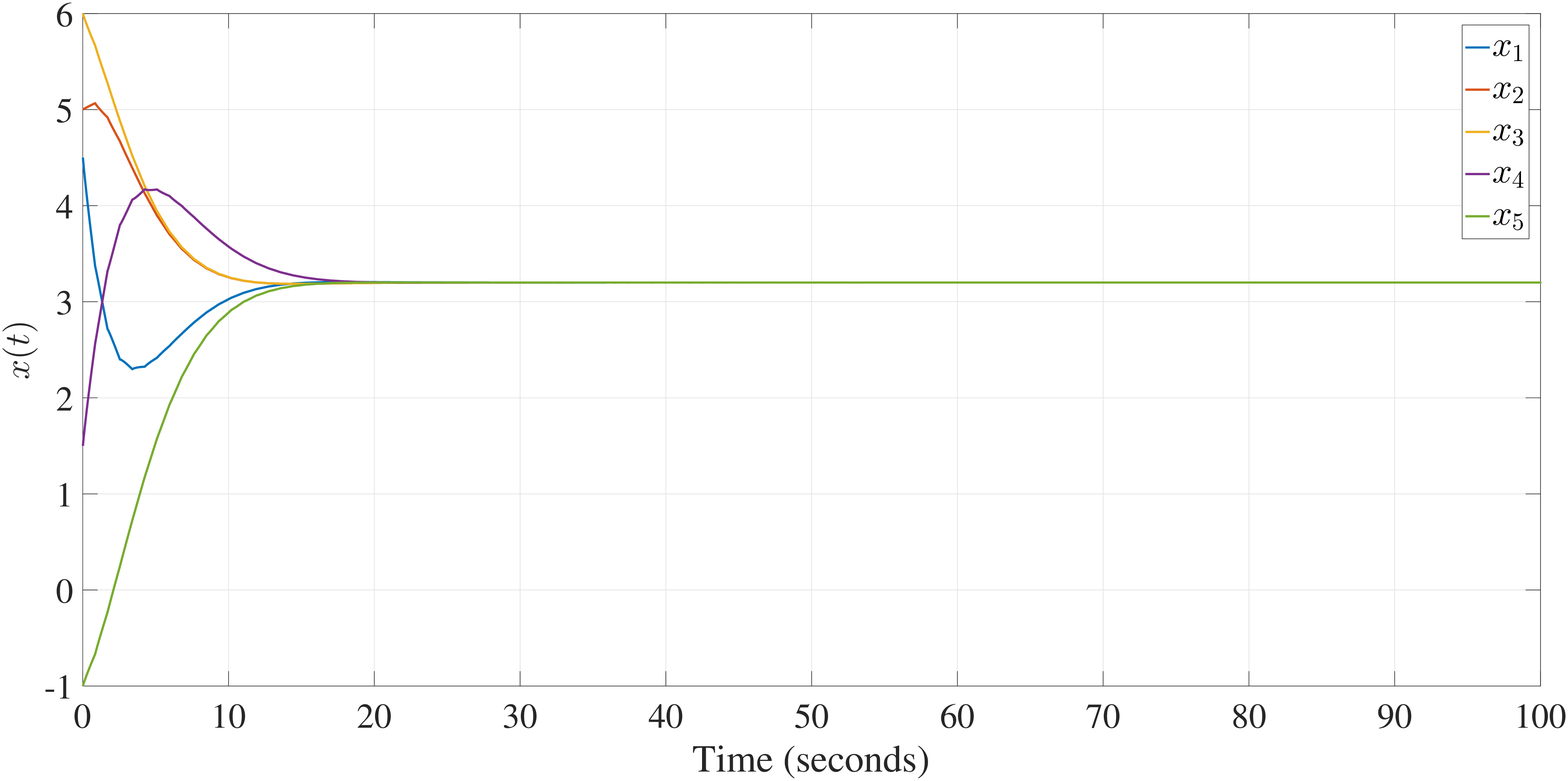}
\caption{The states of (\ref{eq:eqsim}) with the proposed control scheme.}
\label{fig:x}
\end{figure}

\begin{figure}
\centering
\includegraphics[width=10.4cm,height=6.4cm]{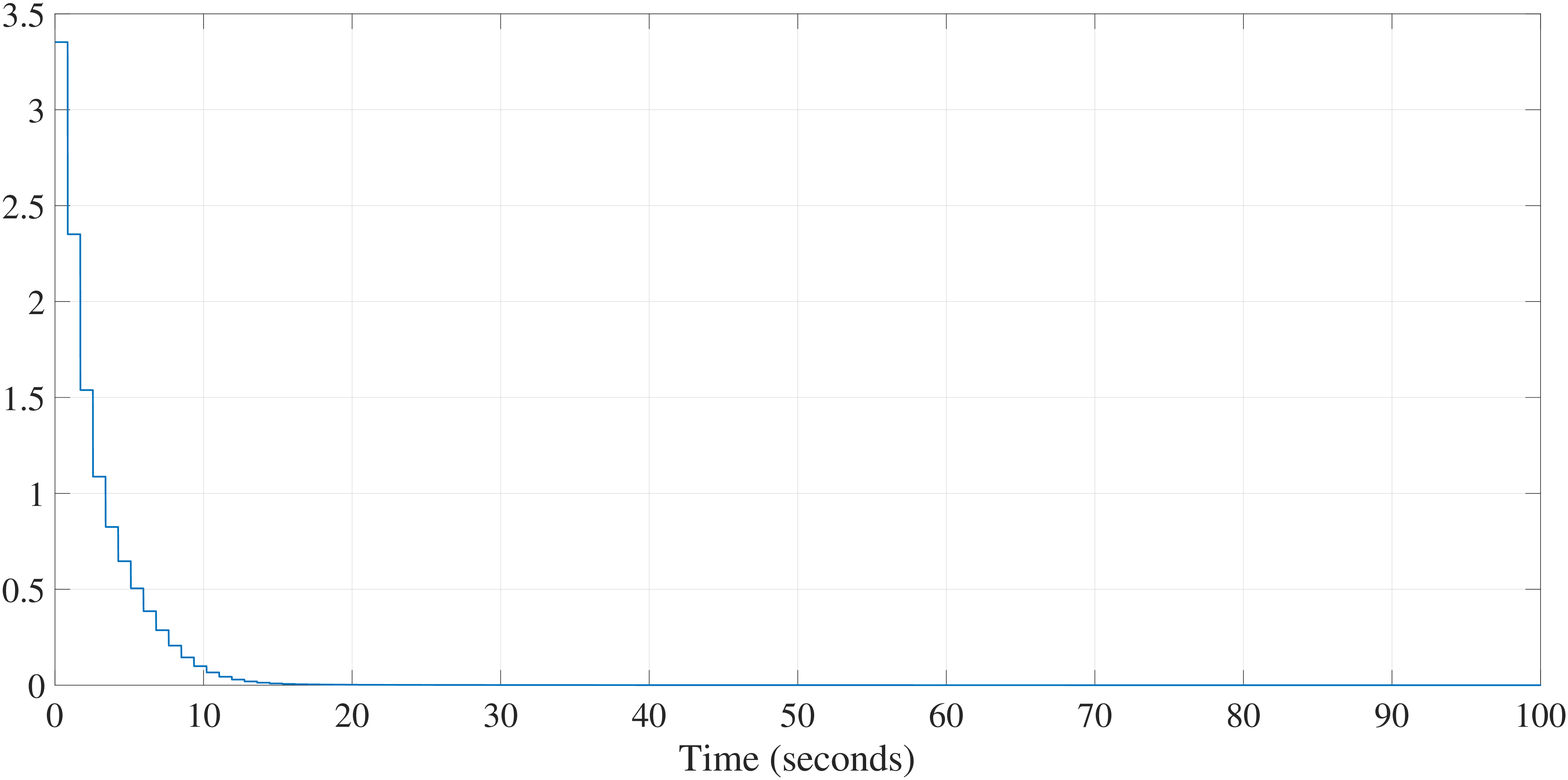}
\caption{The value of $\underline{u}^{\mathrm{T}}\underline{u}$ of the proposed controller.}
\label{fig:u2}
\end{figure}

\begin{figure}
\centering
\includegraphics[width=10.4cm,height=6.4cm]{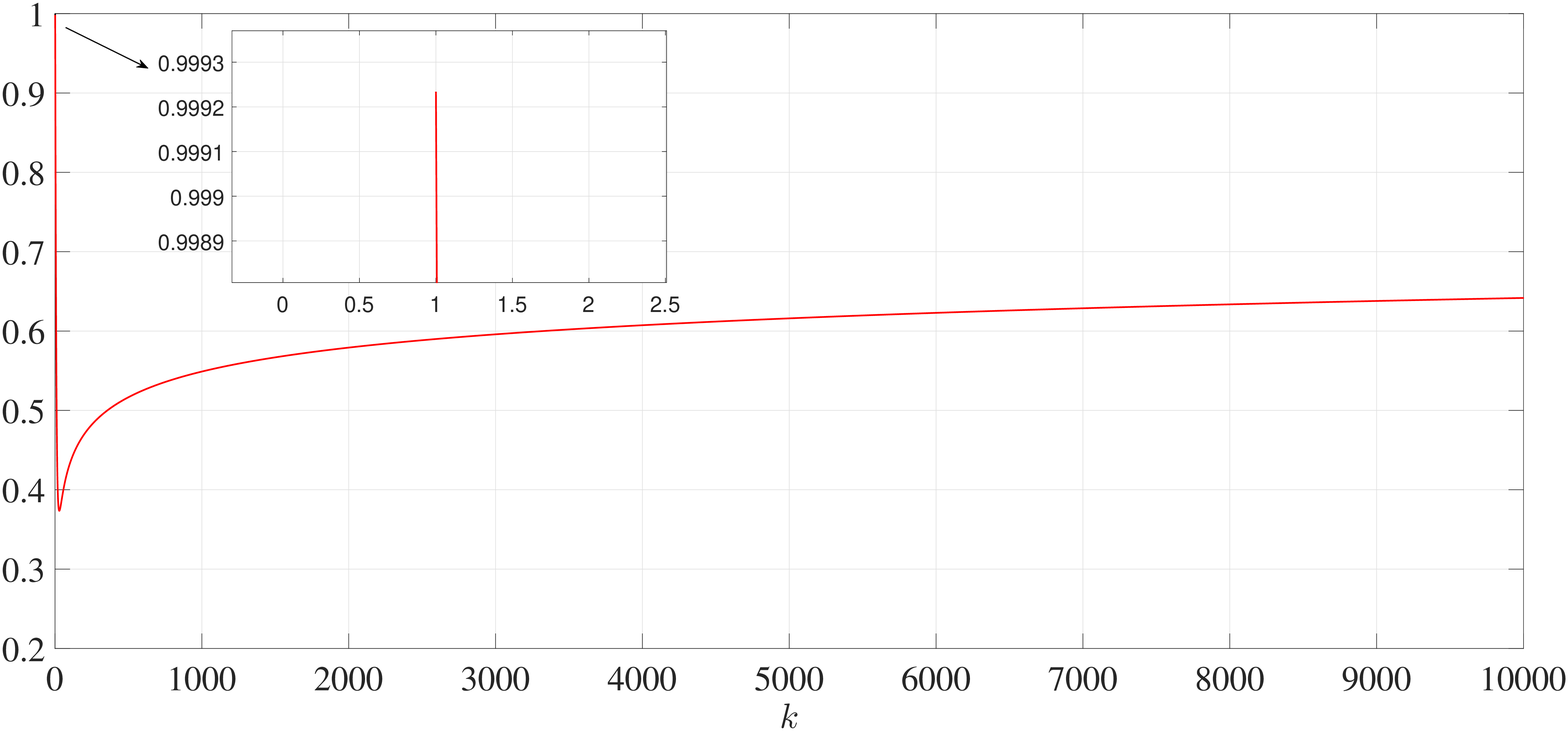}
\caption{The value of $\big[\beta^{k+1}+\sum_{l=1}^{k+1}\lvert f(l)\rvert \big]$ under the proposed control scheme.}
\label{fig:betaf}
\end{figure}

Based on (\ref{eq:eqh}) to (\ref{eq:eqbeta}), the designed control parameter and sampling period are respectively selected as $\gamma =0.15$ and sampling period $h=0.85s$. The simulation results are shown in Fig.~\ref{fig:x} to Fig.~\ref{fig:betaf}. From Fig.~\ref{fig:x}, asymptotic consensus and average-consensus is realized with $\lim_{{t \rightarrow \infty}}x_i(t)=x_{\text{final}}=\frac{\sum_{i=1}^5 x_i(t_0)}{5}=3.2$. The value of $\underline{u}^{\mathrm{T}}\underline{u}$ can be observed from Fig.~\ref{fig:u2}, which shows the boundednesses of all control signals. Furthermore, for the purpose of verifying the condition for ensuring all the control signals are globally bounded, the value of $\big[\beta^{k+1}+\sum_{l=1}^{k+1}\lvert f(l)\rvert \big]$ is given in Fig.~\ref{fig:betaf}, from which it can be noticed that inequality (\ref{eq:eqT2}) holds for $\forall k =0,1,\cdots$ with the selected design parameters.

\begin{figure}
\centering
\includegraphics[width=10.4cm,height=6.4cm]{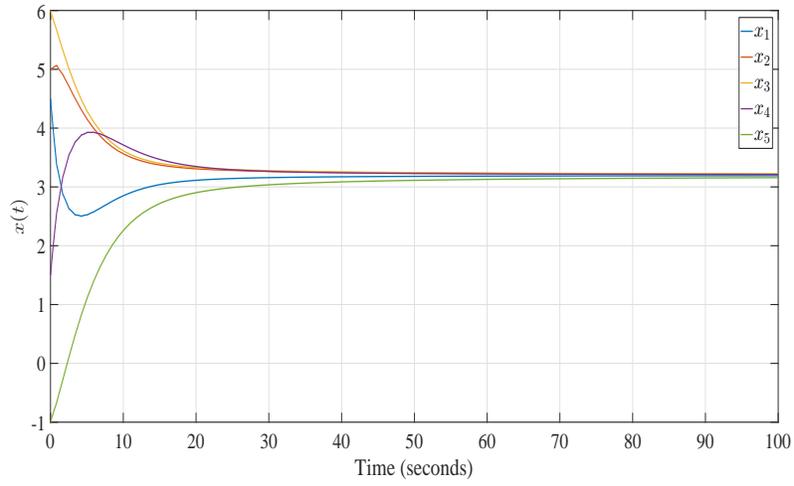}
\caption{The states of (\ref{eq:eqsim}) with control scheme in \cite{yu2017necessary}.}
\label{fig:xxxx}
\end{figure}

\begin{figure}
\centering
\includegraphics[width=10.4cm,height=6.4cm]{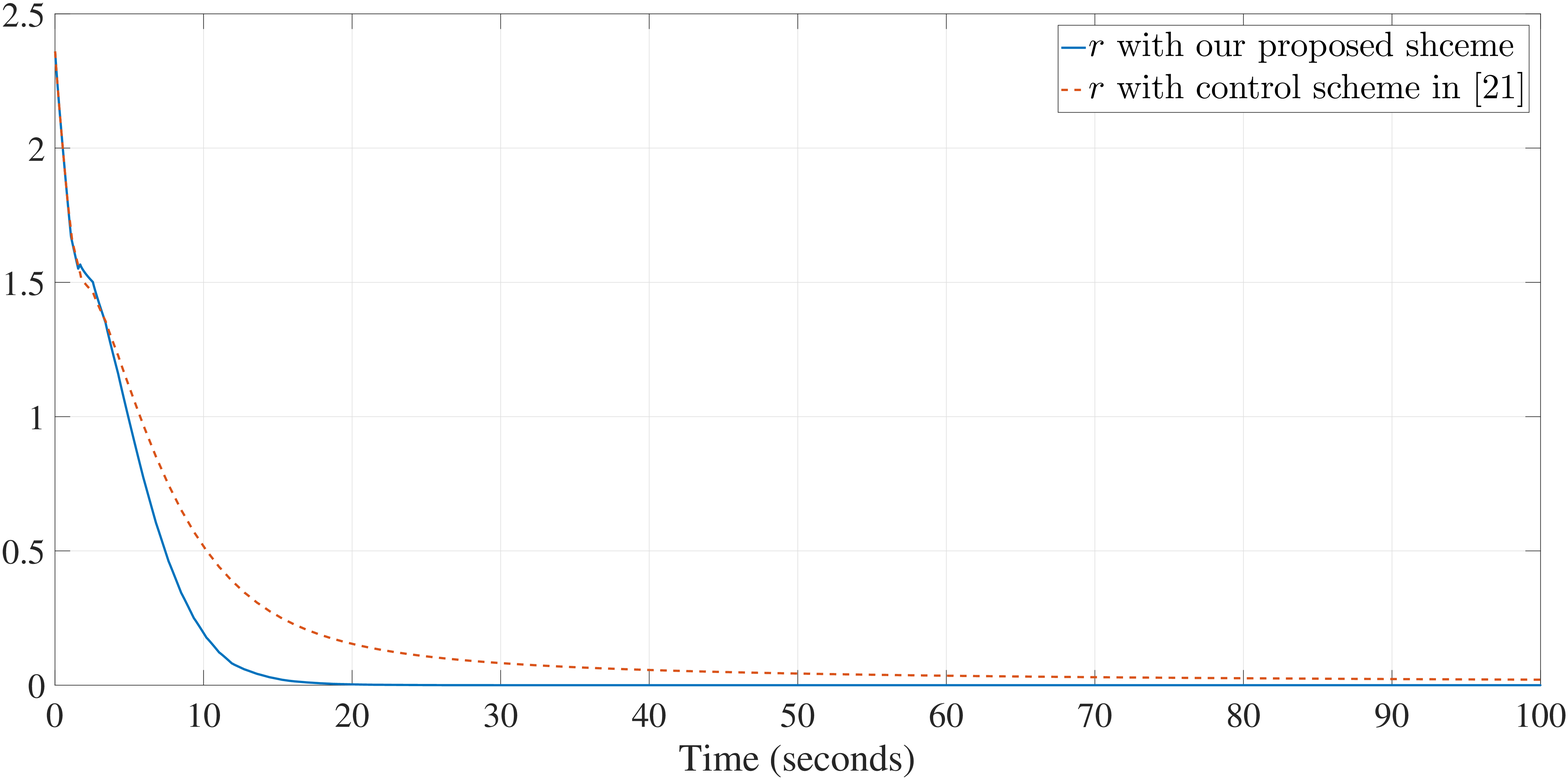}
\caption{The mean absolute error $r$ with proposed scheme and scheme in \cite{yu2017necessary}.}
\label{fig:r}
\end{figure}

\begin{figure}
\centering
\includegraphics[width=10.4cm,height=6.4cm]{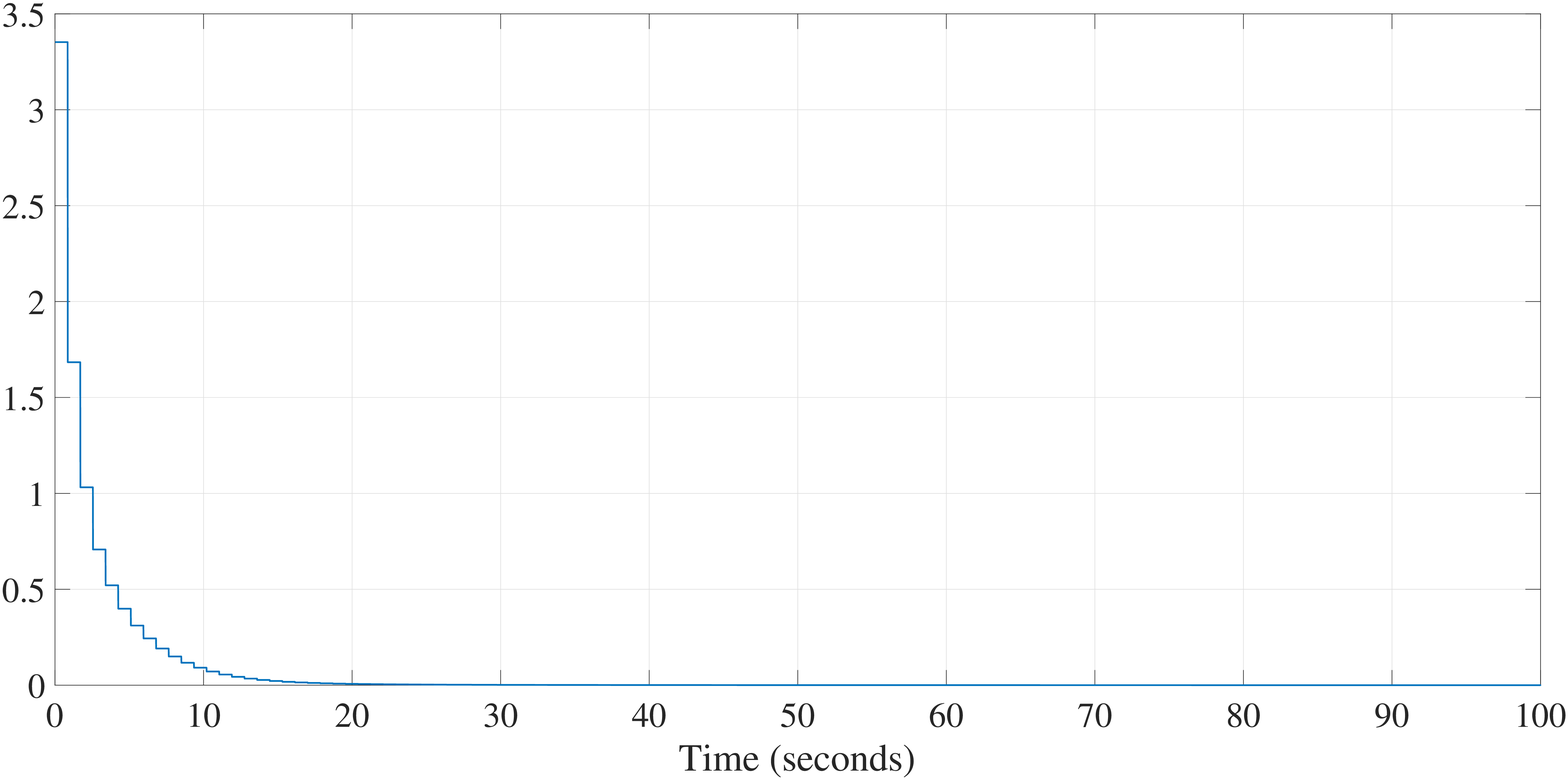}
\caption{The value of $\underline{u}^{\mathrm{T}}\underline{u}$ of the controller in \cite{yu2017necessary}.}
\label{fig:u22}
\end{figure}

To better illustrate the accuracy and effectiveness of our proposed control algorithm, a comparative simulation study between the control scheme in \cite{yu2017necessary} and in this paper is implemented under the same control parameters. 
The mean absolute error $r(t)=\frac{1}{5}\sum_{i=1}^5\lvert x_i(t) - x_{\text{final}}\rvert$ with the proposed scheme and control method in \cite{yu2017necessary} are displayed in Fig.~\ref{fig:r}. Although the control signals under the control scheme in \cite{yu2017necessary} share similar magnitude with our proposed control inputs, which can be observed in Fig.~\ref{fig:u22}, it can be seen from Fig.~\ref{fig:xxxx} and Fig.~\ref{fig:r} that the mean absolute error with control scheme in \cite{yu2017necessary} is larger compared to that with the control scheme in this paper.

\section{Conclusion}
\label{sec:conclusion}
In this paper, a distributed consensus sampled-data based control scheme for multi-agent systems with fractional-order $\alpha \in (0,1)$ is proposed. By taking the hereditary and infinite memory properties of fractional-order calculus into account, a new control scheme is designed to not only ensure system stability, but also achieve asymptotic consensus. Simulation results also illustrate the accuracy and effectiveness of the proposed control algorithm.

\bibliographystyle{unsrt}  

\bibliography{references}

\end{document}